\documentclass[epj]{svjour}

\usepackage{epsfig,amsmath} 
\usepackage{hepnames,hepunits}

\def\cascade{{\sc Cascade}}
\def\pythia{{\sc Pythia}}
\def\herwig{{\sc Herwig}}

\def\mcatnlo{{MC@NLO}}

\def\lsim{\mathrel{\rlap{\lower4pt\hbox{\hskip1pt$\sim$}}
    \raise1pt\hbox{$<$}}}                
\def\gsim{\mathrel{\rlap{\lower4pt\hbox{\hskip1pt$\sim$}}
    \raise1pt\hbox{$>$}}}                

\def\kt{\ensuremath{k_{\rm T}}}
\def\ktz{\ensuremath{k_{\rm T,0}}}
\def\pt{\ensuremath{p_{\rm T}}}
\def\qt{\ensuremath{q_{\rm t}}}
\def\mdy{\ensuremath{m_{\rm DY}}}

\newcommand{\Pmax}{\mu^2}

\newcommand{\alphas}{\ensuremath{\alpha_\mathrm{s}}}

\newcommand{\PBM}{PB}

\newenvironment{tolerant}[1]{\par\tolerance=#1\relax}{ \par }

\newcommand{\dglap}{Gribov:1972ri,Lipatov:1974qm,Altarelli:1977zs,Dokshitzer:1977sg}

\begin{document}

\title{\boldmath  The transverse momentum spectrum of low mass Drell-Yan production at next-to-leading order in the parton branching method}

\author{A.~Bermudez~Martinez\inst{1} \and 
P.L.S.~Connor \inst{1} \and 
D.~Dominguez~Damiani \inst{1} \and 
L.I.~Estevez~Banos \inst{1} \and 
F.~Hautmann \inst{2,3}  \and
H.~Jung \inst{1} \and 
J.~Lidrych \inst{1} \and 
A.~Lelek\inst{3} \and 
M.~Mendizabal \inst{1} \and 
M.~Schmitz \inst{1} \and 
S.~Taheri~Monfared \inst{1} \and 
Q.~Wang \inst{1,4} \and 
T.~Wening \inst{1} \and 
H.~Yang  \inst{1,4} \and 
R.~\v{Z}leb\v{c}\'{i}k \inst{1}}

\institute{Deutsches Elektronen-Synchrotron, D-22607 Hamburg,
\and RAL, Chilton OX11 0QX and University of Oxford, OX1 3NP,
\and Elementaire Deeltjes Fysica, Universiteit Antwerpen, B 2020 Antwerpen, Belgium,
\and School of Physics, Peking University}


\abstract{
It has been observed in the literature that measurements of low-mass Drell-Yan (DY) transverse 
momentum spectra at low center-of-mass energies $\sqrt{s}$ are not well described by perturbative 
QCD calculations in collinear factorization in the region where transverse momenta are comparable 
with the DY mass. We examine this issue from the standpoint of the Parton Branching (PB) 
method, combining next-to-leading-order (NLO) calculations of the hard process with the evolution 
of transverse momentum dependent (TMD) parton distributions. 
We compare our predictions with experimental measurements at low DY mass, and find very good agreement.
In addition we use the low mass DY measurements at low $\sqrt{s}$   to determine the width $q_s$ of the intrinsic Gauss distribution of the
\PBM -TMDs at low evolution scales.  We  find values close to what has earlier been used in applications of  \PBM -TMDs to 
high-energy processes  at the   Large Hadron Collider (LHC) and HERA. 
We find that at low DY mass and low $\sqrt{s}$ even in the region of $\pt/\mdy \sim 1$ the contribution 
of multiple soft gluon emissions (included in the \PBM -TMDs) is essential to describe the measurements, while at larger masses  
($\mdy \sim m_{\PZ }$) and  LHC energies the contribution from soft gluons in the region of  $\pt/\mdy \sim 1$ is small.
}


\titlerunning{The $\pt$ spectrum of low mass DY production at NLO order in the PB method}
\maketitle
\flushbottom

\section{Introduction} 
\label{Intro}

Higher-order perturbative QCD calculations are required for a precise description of Drell-Yan (DY) production~\cite{Drell:1970wh}  measurements in $\Pp\Pp$ collisions at 
the LHC~\cite{Khachatryan:2016nbe,Aad:2015auj,Aad:2014qja,Chatrchyan:2011wt,Aad:2019wmn,Sirunyan:2019bzr}. The production of \PZ -bosons at transverse momenta smaller than  the boson mass ($\pt < {\cal O}(m_{\PZ })$) cannot be described by fixed order calculations, but soft gluon resummation to all orders~\cite{Dokshitzer:1978yd,Parisi:1979se,Curci:1979am,Altarelli:1984pt,Collins:1984kg} is needed, as featured in various  analytical TMD resummation methods~\cite{Bizon:2018foh,Bizon:2019zgf,Catani:2015vma,Scimemi:2017etj,Bacchetta:2019tcu,Bacchetta:2018lna,Ladinsky:1993zn,Balazs:1997xd,Landry:2002ix,resbosweb,Alioli:2015toa,Bozzi:2019vnl,Baranov:2014ewa}  or in parton showers 
of multi-purpose Monte Carlo (MC) event generators \cite{Sjostrand:2014zea,Bellm:2015jjp,Bahr:2008pv,Gleisberg:2008ta} 
matched with higher-order matrix elements \cite{Frixione:2003ei,Frixione:2002ik,Frixione:2007vw,Nason:2012pr,Alwall:2014hca,Frederix:2015eii}. In Ref.~\cite{Martinez:2019mwt} it  was proposed that  the \PZ -boson \pt\ spectrum can be accurately evaluated by  using  the Parton Branching (PB) 
formulation~\cite{Hautmann:2017xtx,Hautmann:2017fcj} of TMD evolution together   with NLO calculations of the hard scattering process in 
the  {\sc MadGraph5\_aMC@NLO} \cite{Alwall:2014hca} framework. The predictions thus obtained were found to be in very good agreement with measurements 
from ATLAS at $\sqrt{s}=8$~\TeV~\cite{Aad:2015auj}  and CMS at $\sqrt{s}=13$~\TeV\cite{Sirunyan:2019bzr}, 
 with modest sensitivity to the non-perturbative (intrinsic-\kt ) part of the TMD distributions~\cite{Angeles-Martinez:2015sea}.

The transverse momentum spectrum of DY production at lower mass $m_{DY}$ allows one to study in more detail the non-perturbative contribution, as the phase space for perturbative evolution is reduced. However,  the measurement of the transverse momentum at low mass of the DY pair is experimentally very challenging, since one has to measure down to low transverse momenta of the decay leptons, where experimental background and misidentification of the DY lepton pairs can be significant. At the LHC the lowest DY mass used for the low transverse momentum spectra ($\pt$  
{\ \hbox{\raise 2pt \hbox{$>$} \kern -13pt
                     \lower 3pt \hbox{$\sim$}} 
 1 \GeV) is $\sim 46$~\GeV ~\cite{Aad:2015auj}, while at lower center-of-mass energies DY  measurements covering the low \pt\ region  for lower masses exist 
from PHENIX~\cite{Aidala:2018ajl} at $\sqrt{s}=200$~\GeV , from R209~\cite{Antreasyan:1981eg} at $\sqrt{s}=62$~\GeV , and  from NuSea~\cite{Webb:2003ps,Webb:2003bj} and E605~\cite{Moreno:1990sf} at  $\sqrt{s}=38.8$~\GeV . 
In a study~\cite{Gieseke:2007ad} based on the Monte Carlo event generator \herwig , good agreement with measurements at low $\sqrt{s}$ was found 
after changing parameters for the parton shower and intrinsic transverse momentum.
The description of these measurements is discussed in terms of TMDs in Refs.~\cite{Bacchetta:2019tcu,Scimemi:2019cmh,Bacchetta:2019sam,Hautmann:2020cyp}. 
In Ref.~\cite{Bacchetta:2019tcu} these measurements were compared with collinear NLO predictions and significant discrepancies were observed. 

In this paper we apply the TMD parton densities obtained using the PB method (fitted~\cite{Martinez:2018jxt} to  inclusive deep-inelastic scattering  (DIS) precision data from HERA) together with an NLO calculation of DY production \cite{Alwall:2014hca} precisely in the same manner as in Ref.~\cite{Martinez:2019mwt}, but now to treat low-mass DY production. We first briefly review the main elements of the PB approach and the matching of the PB-TMDs with the NLO calculation (Sec.~\ref{sec:aMCatNLO}). 
Then we show that these low energy  measurements are very well described with the PB-MCatNLO approach in the whole region of $\pt/\mdy$  
(in contrast to the observation in Ref.~\cite{Bacchetta:2019tcu}) and examine  the role  of both the perturbative evolution and  
the non-perturbative (intrinsic-\kt ) distribution  (Sec.~\ref{sec:lowmass}).  
We  provide a discussion to put these results in a  broader context (Sec.~\ref{sec:disc}), and   
  finally give conclusions (Sec.~\ref{sec:concl}). 
 
 \section{ \PBM -TMDs  and DY production at NLO }
\label{sec:aMCatNLO}

In this section we recall the basic elements of the PB approach, and illustrate the main features of applying it 
to DY production at different center-of-mass energies, from fixed-target experiments to  the LHC.  

\subsection{Collinear and TMD densities from the PB method}
\label{PBTMD}
The approach proposed in~\cite{Hautmann:2017xtx} allows 
evolution equations for  both collinear  and TMD parton distributions to be 
solved numerically with the \PBM\ method. In this approach, the concept 
of resolvable and non-resolvable branchings is applied by using Sudakov form factors. A soft-gluon resolution scale $z_M$ is 
introduced to separate   resolvable and non-resolvable branchings. The Sudakov form factors, which describe the evolution without resolvable branching 
 from one scale $\mu_0$ to another scale $\mu$, are given in terms of the resolvable splitting probabilities  $P_{ba}^{(R)} (\alphas , z ) $ as follows, 
\begin{eqnarray}
\label{sud-def}
  \Delta_a ( z_M, \mu^2 , \mu^2_0 )   
&  = &\exp \left(  -  \sum_b  \int^{\mu^2}_{\mu^2_0} 
{{d \mu^{\prime 2} } 
\over \mu^{\prime 2} } \right.  \\
&& \left. \int_0^{z_M} dz \  z 
\ P_{ba}^{(R)}\left(\alphas , 
 z \right)   \nonumber
\right) 
  \;\; ,   
\end{eqnarray}
where 
$a , b$ are flavor indices,  $\alphas$ is the strong coupling,   $z$ is the longitudinal momentum splitting variable, and   $z_M < 1 $ is the soft-gluon resolution parameter. A detailed description of the  \PBM\ method is given  in Refs.~\cite{Martinez:2018jxt,Hautmann:2017fcj}.

\begin{figure}[h!tb]
\begin{center} 
\includegraphics[width=0.24\textwidth]{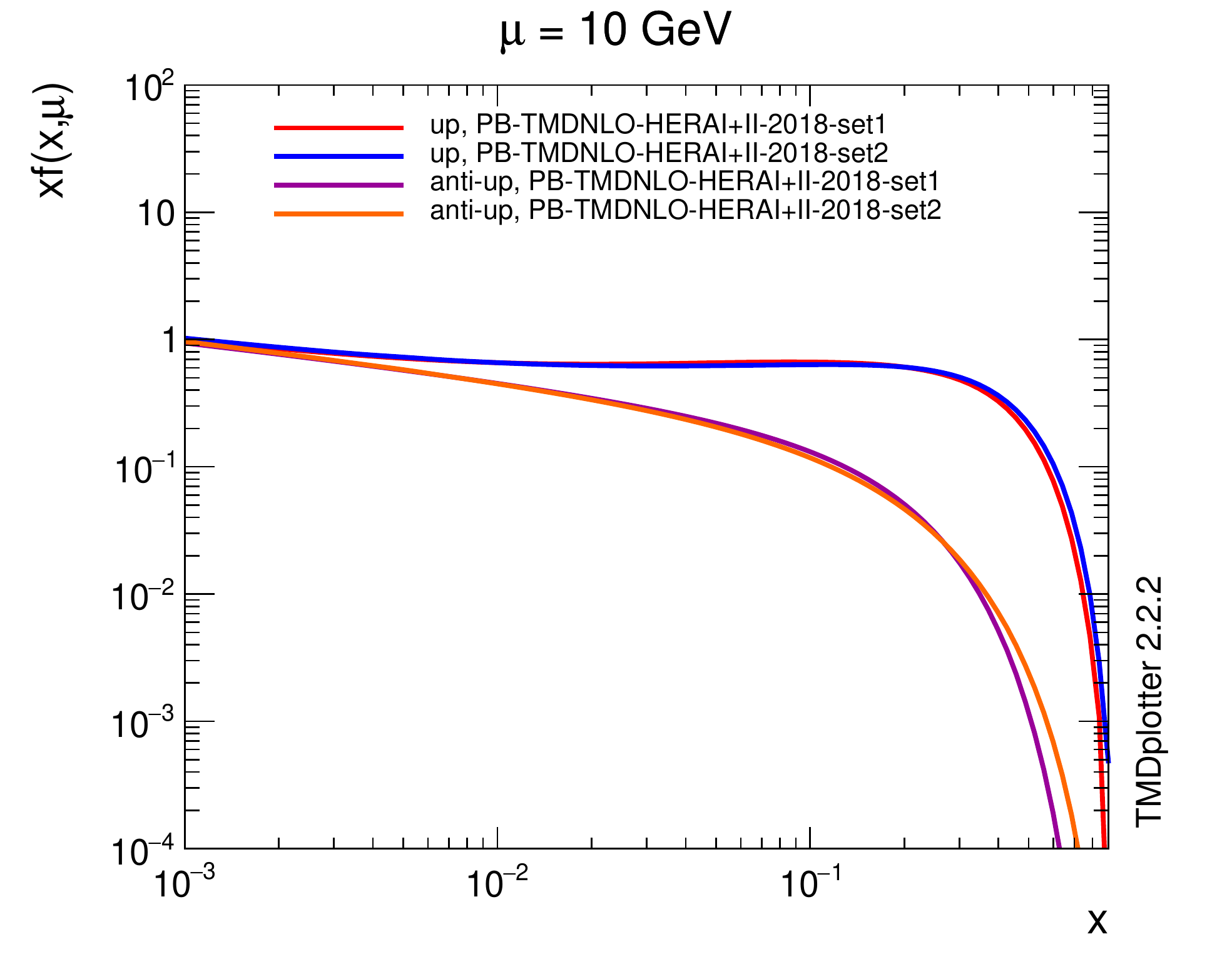} 
\includegraphics[width=0.24\textwidth]{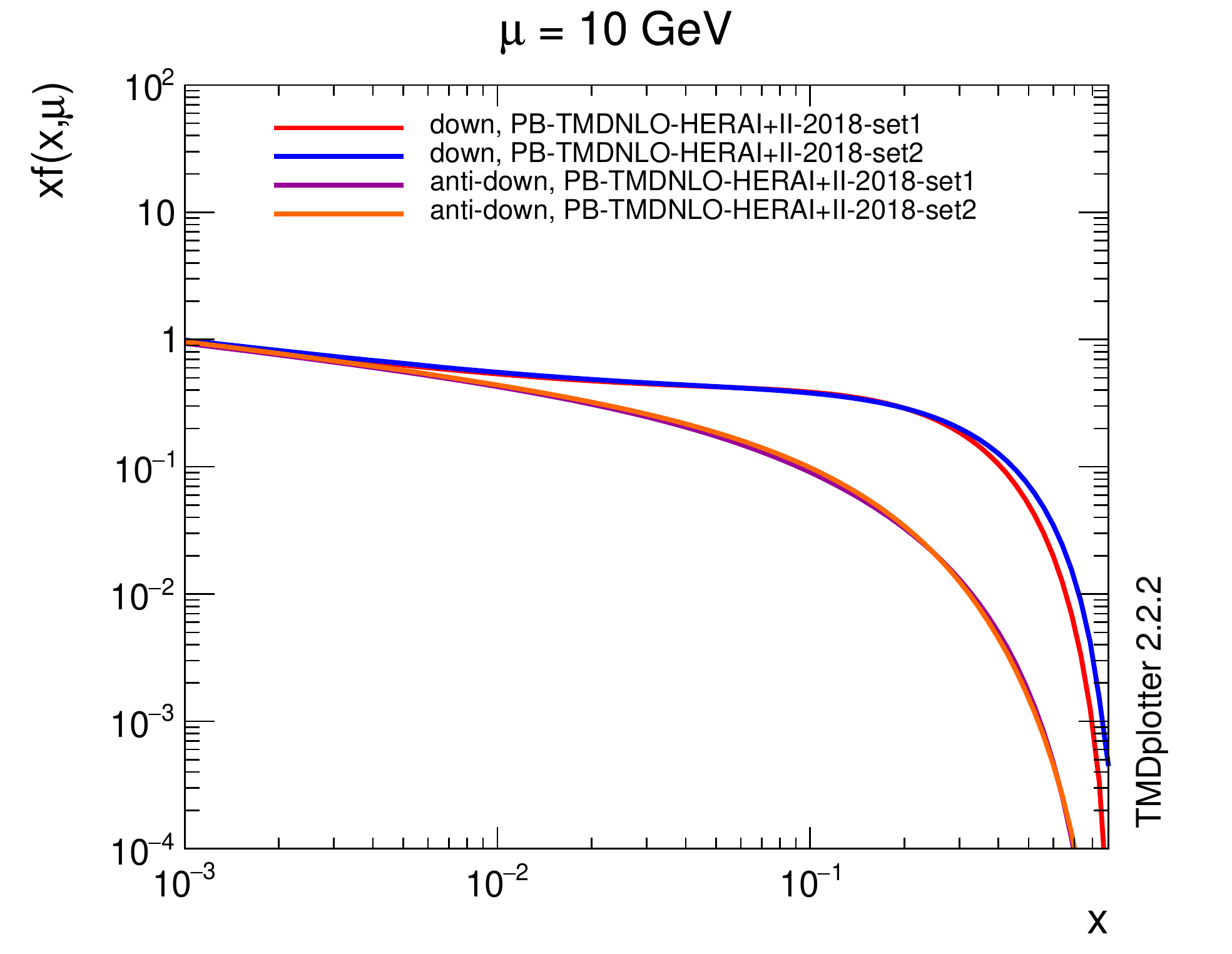} 
\includegraphics[width=0.24\textwidth]{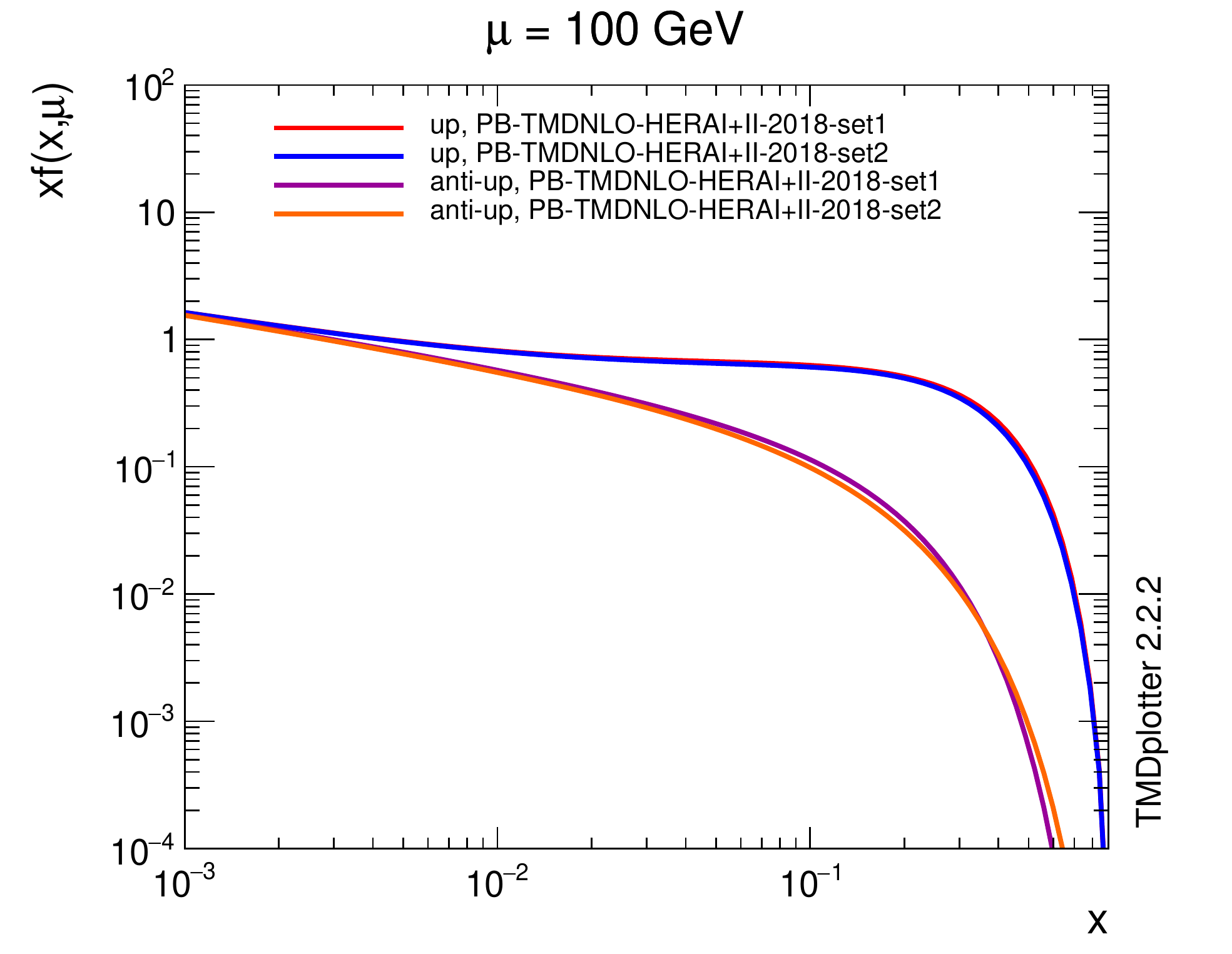} 
\includegraphics[width=0.24\textwidth]{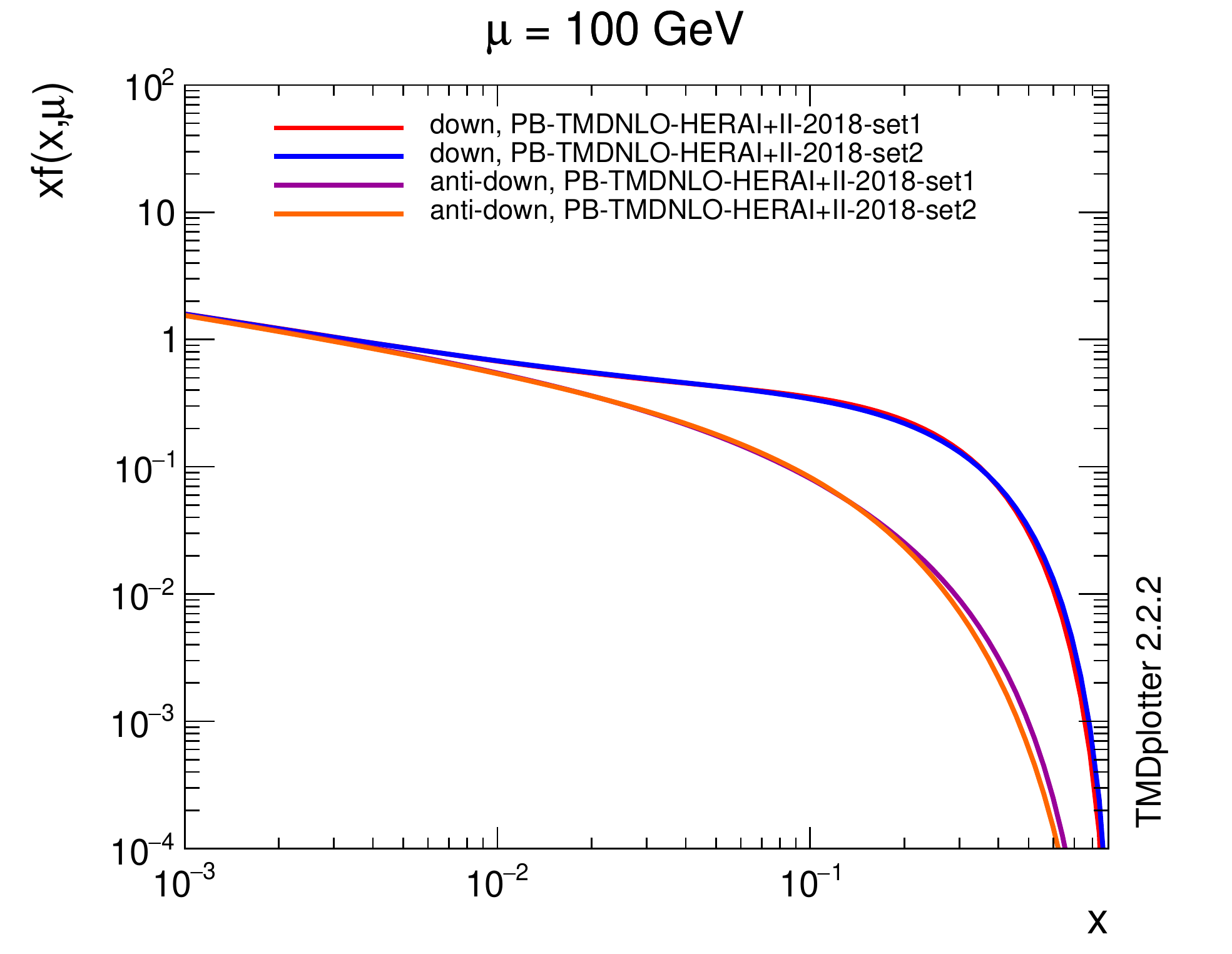} 
  \caption{\small Collinear parton distributions for up and down quarks  (PB-NLO-2018-Set1, PB-NLO-2018-Set 2) as a function of $x$ at   $\mu = 10 $ and $100$ \GeV.
  }
\label{TMD_pdfs1}
\end{center}
\end{figure} 

 \begin{tolerant}{3500}
The TMD parton density distributions are obtained from the non-perturbative starting distributions  ${\cal A}_{0,b} (x',\ktz^2,\mu_0^2)$ after convoluting with a perturbative evolution kernel $ {\cal  K}_{ba}$.
As described in~\cite{Martinez:2018jxt},  we have 
\end{tolerant}
\begin{eqnarray}
x{\cal A}_a(x,\kt^2,\mu^2) 
 &= &
 x\int dx' \int dx'' {\cal A}_{0,b} (x',\ktz^2,\mu_0^2)  \nonumber \\
  & & {\cal  K}_{ba}\left(x'',\ktz^2,\kt^2,\mu_0^2,\Pmax\right) 
 \delta(x' x'' - x) 
\nonumber  
\\
& = & \int dx' {\cal A}_{0,b} (x',\ktz^2,\mu_0^2)
\frac{x}{x'} \nonumber  \\
& &
{ {\cal  K}_{ba}\left(\frac{x}{x'},\ktz^2,\kt^2,\mu_0^2,\Pmax\right) }  \;\; .
\label{TMD_kernel}
\end{eqnarray}
We use a factorized form for the starting distribution ${\cal A}_0$, for simplicity, (in general, the $\ktz$ distribution can be also flavor and $x$-dependent),
\begin{eqnarray}
{\cal A}_{0,b} (x,\ktz^2,\mu_0^2)  & = & f_{0,b} (x,\mu_0^2) \nonumber \\ 
&& \cdot \exp\left(-| \ktz^2 | / 2 \sigma^2\right) / ( 2 \pi \sigma^2) \; , 
\label{TMD_A0}
\end{eqnarray}
with $ \sigma^2  =  q_s^2 / 2 $  independent of the parton flavor  and $x$,  with a constant value $q_s = 0.5$~\GeV. 
Also, the evolution kernels $ {\cal  K}_{ba}$ in Eq.~(\ref{TMD_kernel}) do not include  any non-perturbative component. In principle, 
non-perturbative contributions to Sudakov form factors could be introduced in the $ {\cal  K}_{ba}$ kernels of the PB method,  
and parameterized in terms of non-perturbative functions to be fitted to 
experimental data (similarly to what is done in other approaches, e.g.~\cite{Ladinsky:1993zn,Balazs:1997xd,Landry:2002ix,resbosweb}). 
  For simplicity, however, at present we take the kernels $ {\cal  K}_{ba}$  to be purely perturbative. 
\begin{figure}[h!tb]
\begin{center} 
\includegraphics[width=0.24\textwidth]{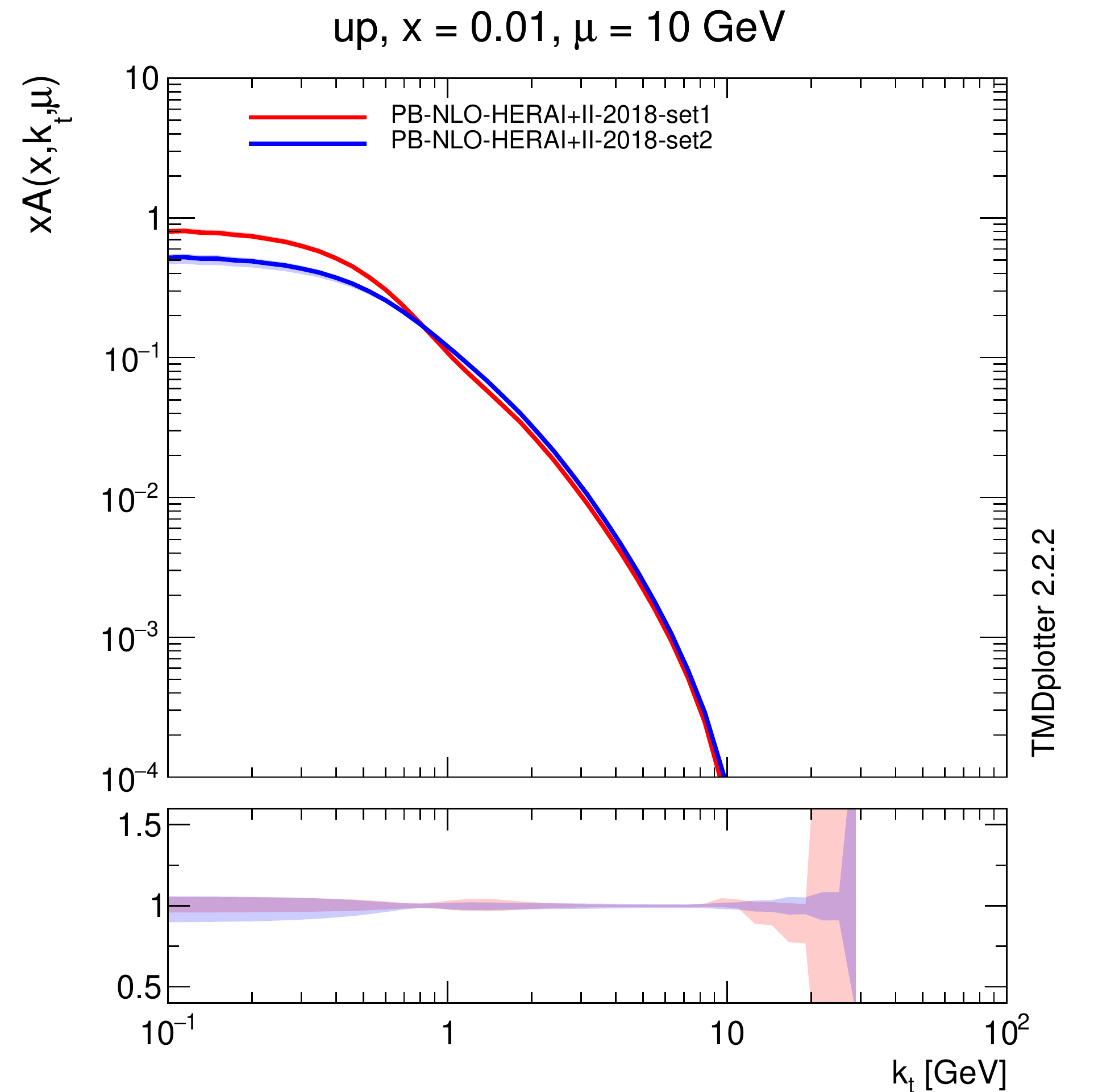} 
\includegraphics[width=0.24\textwidth]{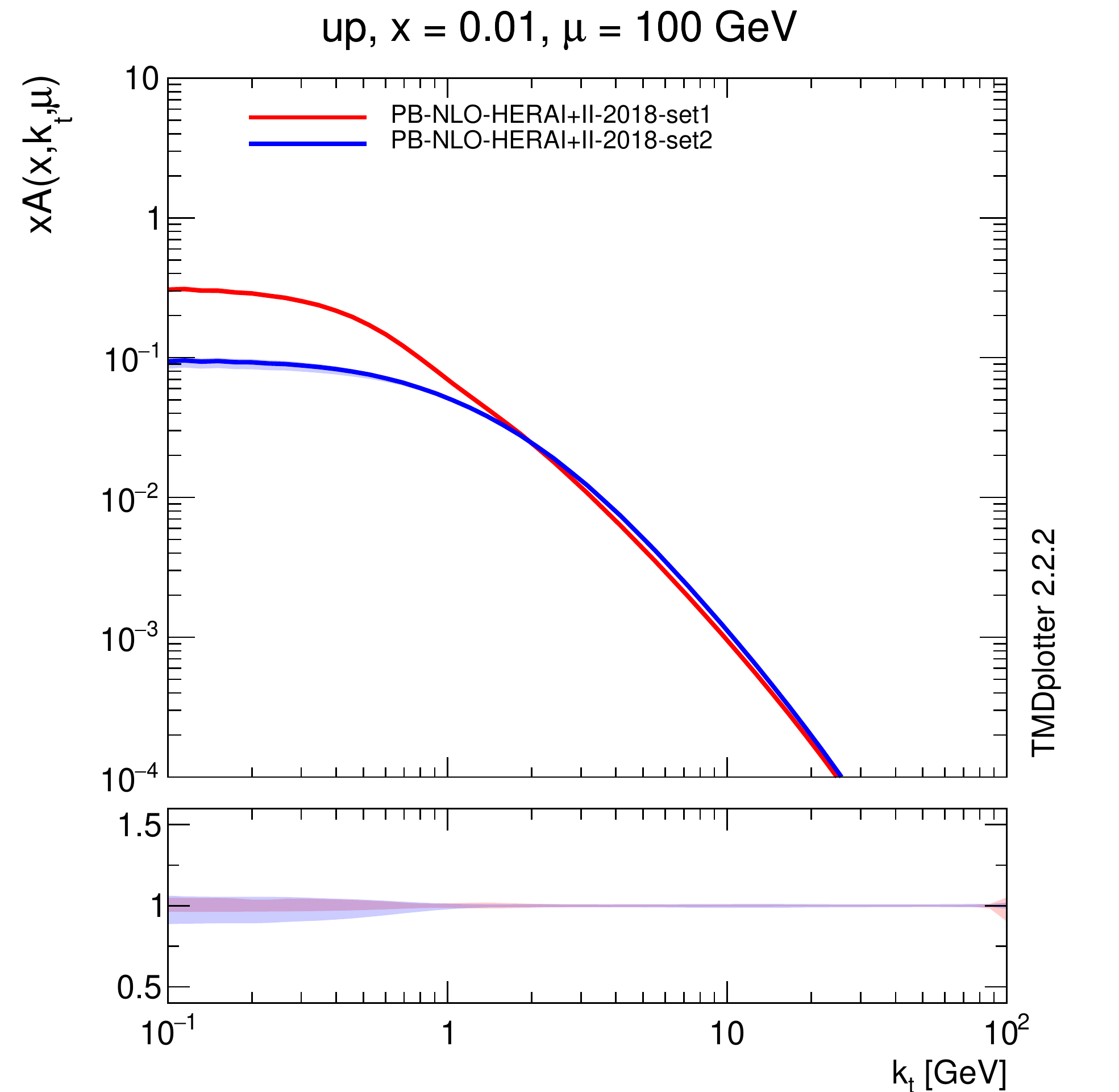} 
  \caption{\small TMD parton distributions for up quarks  (PB-NLO-2018-Set1 and PB-NLO-2018-Set 2) as a function of $\kt$ at $\mu=10$ and $100$  \GeV\  and $x=0.01$.
  In the lower panels  the full uncertainty of the TMDs is shown, as obtained from the fits \protect\cite{Martinez:2018jxt}.
  }
\label{TMD_pdfs2}
\end{center}
\end{figure} 

The \PBM\ method enables the explicit calculation of the kinematics at every branching vertex, once the evolution scale is specified in terms of kinematic variables. 
In Ref.~\cite{Hautmann:2017xtx} it was pointed out that  angular ordering  gives 
 transverse momentum distributions which are stable with respect to variations of the resolution parameter  $z_M$.
 In angular ordering, the angles of the emitted partons increase from the hadron side towards the hard scattering~\cite{Marchesini:1987cf,Catani:1990rr}.
The transverse momentum of the $i$'s emitted parton $q_{t\,i}$ can be calculated in terms of the angle $\Theta_i$ of the emitted parton with respect to 
 the beam directions from $q_{t,i} = (1-z_i) E_{i} \sin \Theta_i$. Associating the "angle"  $E_i \sin \Theta_i$ with $\mu_i$ gives 
\begin{equation}
  \label{ang-ordering}
 {q}_{t,i}^2  =  (1-z_i)^2 \mu_i^ 2  \;\; .
\end{equation}

Deep-inelastic scattering measurements from HERA are used in Ref. \cite{Martinez:2018jxt}  to determine the free parameters of the  starting distributions  at scale $\mu_0 \sim $ 1 GeV.
The fits were performed 
using the  open-source fitting platform   \verb+xFitter+~\cite{Alekhin:2014irh} and  a new development described in Ref.~\cite{Hautmann:2017fcj,Martinez:2018jxt} of the  
numerical techniques~\cite{Hautmann:2014uua}.
Collinear and  TMD distributions were extracted including the determination of experimental and theoretical uncertainties.
In Ref. \cite{Martinez:2018jxt} two sets of parton distributions are described: Set~1, which uses the evolution scale as argument  in the running coupling $\alphas$, similar to what is used in HERAPDF 2.0 NLO~\cite{Abramowicz:2015mha}, and Set~2, which uses the transverse momentum in the evolution of $\alphas$.

 For soft gluon resolution      $z_M \to 1 $ and strong coupling $\alphas \to \alphas   (  \mu^{\prime 2} ) $ it was verified~\cite{Martinez:2018jxt,Hautmann:2017fcj} numerically, 
with a numerical accuracy of better than 1 \% over a range of five orders of magnitude  both in $x$ and in $\mu$, 
 that DGLAP evolution equations~\cite{\dglap } are recovered from PB evolution. 

In Fig.~\ref{TMD_pdfs1} the Set~1 and Set~2  collinear  densities are shown for up-quark and down-quark  at  evolution scales of $\mu=10$ and $100$  \GeV . 
In Fig.~\ref{TMD_pdfs2}   we show the TMD distributions for up-quarks at $x=0.01$ and $\mu=10$ and $100$ \GeV .
The plots in Figs.~\ref{TMD_pdfs1},\ref{TMD_pdfs2}   are made using the TMDplotter tool~\cite{Hautmann:2014kza,Connor:2016bmt}.
Collinear densities are given in a format compatible with LHAPDF~\cite{Buckley:2014ana}.

 \subsection{Matching \PBM -TMDs  with NLO calculations}
\label{matching}

We employ the approach proposed in Ref.~\cite{Martinez:2019mwt} to perform the matching of 
 \PBM -TMDs  with the NLO calculation of DY production. In this subsection  we briefly describe a few technical aspects of the computation and analyze 
 numerically the   contributions of  \PBM -TMDs  and  NLO in the matching procedure. 

\begin{tolerant}{1200}
Following \cite{Martinez:2019mwt},  {\sc MadGraph5\_aMC@NLO} (version 2.6.4, hereafter  labelled \mcatnlo )  \cite{Alwall:2014hca}  is used to calculate the Drell-Yan process 
at NLO, i.e.,  including ${\cal O }  (\alpha_s) $ corrections to the hard-scattering matrix element, 
 together with the NLO \PBM\ parton distributions (Set~2) of Ref.~\cite{Martinez:2018jxt}.
As in~\cite{Martinez:2019mwt}, motivated by the angular ordering in PB evolution, we use \herwig 6 \cite{Corcella:2002jc,Marchesini:1991ch} subtraction terms in \mcatnlo . 
A similar method to describe DY production at leading order is proposed in Ref.~\cite{Golec-Biernat:2019scr}.   
\end{tolerant}

A matching scale $\mu_{m}$ (parameter \verb+SCALUP+) separates the contribution of the real emission treated by the matrix element calculation and the contribution from the \PBM -TMD.

\begin{figure}[h!tb]
\begin{center} 
\includegraphics[width=0.45\textwidth]{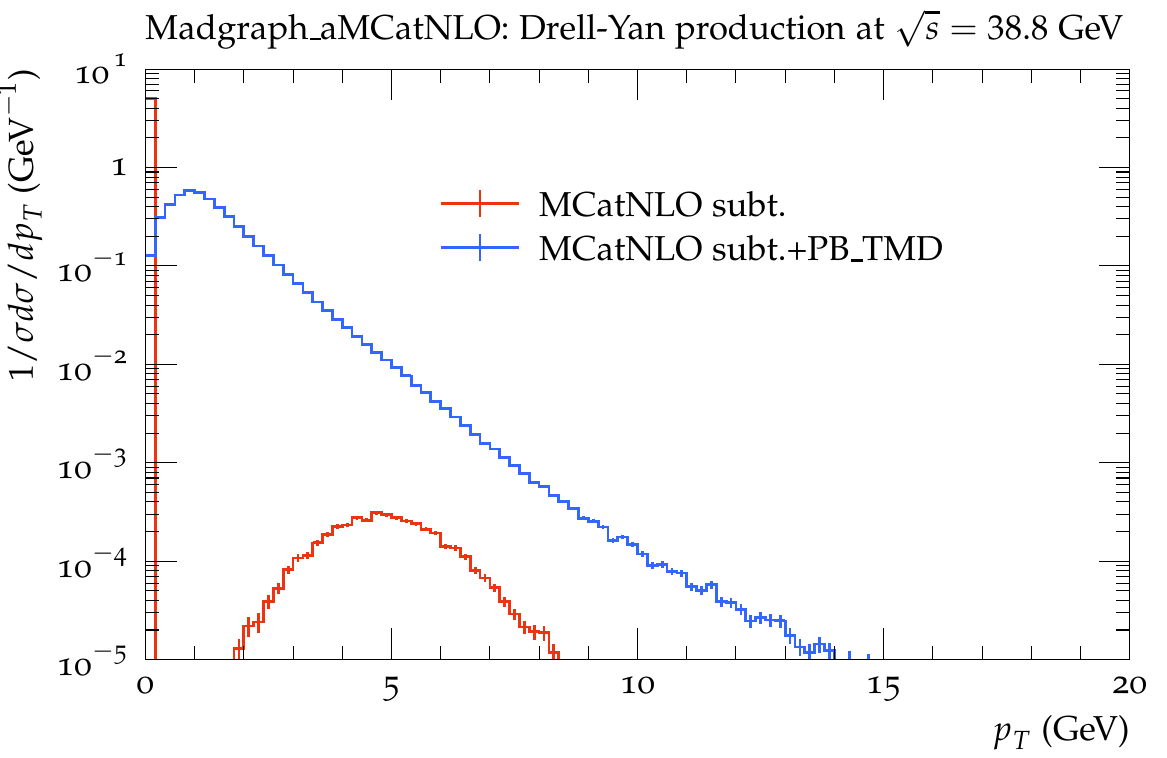} 
\includegraphics[width=0.45\textwidth]{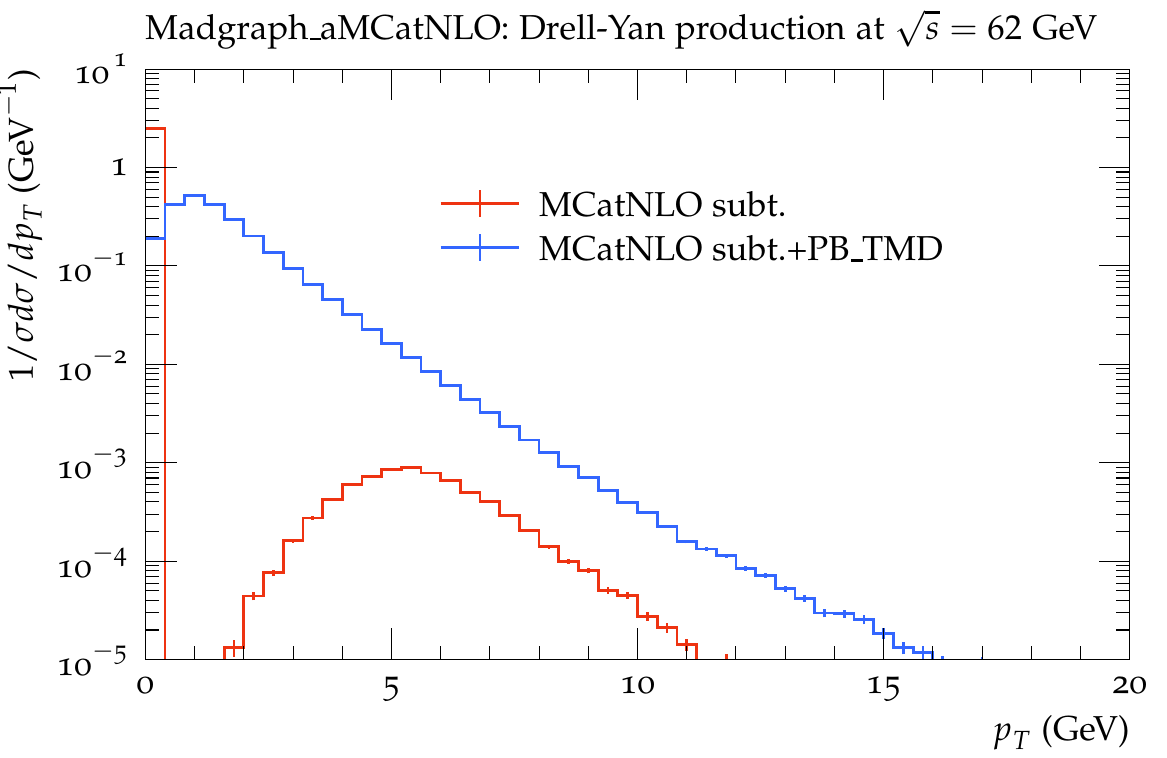} 
\includegraphics[width=0.45\textwidth]{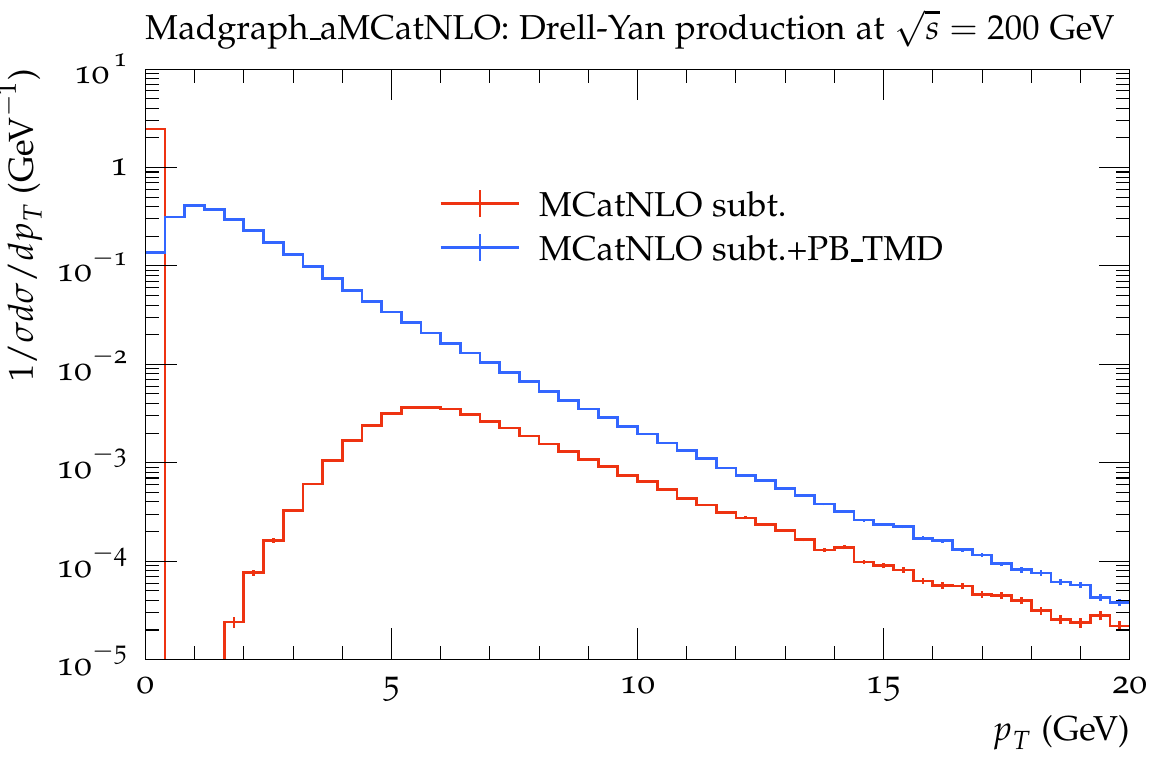} 
  \caption{\small Transverse momentum spectrum of DY production  at parton level (LHE level) for  subtraction terms and after inclusion of \PBM -TMDs. Distributions are shown for $\mdy > 4$~\GeV\  at $\sqrt{s}=38.8$~\GeV , at  $\sqrt{s}=62$~\GeV\ and  at $\sqrt{s}=200$~\GeV .
 }
\label{mcatnlo_lhe}
\end{center}
\end{figure}

The hard process is calculated at a scale $\mu$ and the longitudinal momentum fraction $x$, 
where 
  $\mu = \frac{1}{2} \sum_i \sqrt{m^2_i +p^2 _{t,i}}$, with the sum running over the decay products and the final jet. The same scales are used in the \PBM -TMD. The scale $\mu_{m}=$\verb+SCALUP+ is also used as an upper limit for the transverse momentum (the calculation are performed with the \cascade 3 package ~\cite{Jung:2010si,Jung:2001hx} (version  \verb+3.0.X+)). We employ Rivet~\cite{Buckley:2010ar} to analyze output files.

In Fig.~\ref{mcatnlo_lhe} we show results,  at different   center-of-mass energies $\sqrt{s}$,   
for   the  DY lepton-pair  transverse momentum distribution,  
obtained from the  \mcatnlo\   calculation at a purely partonic level (LHE level) 
 using  \herwig 6 subtraction terms   (red solid curves in the plots), and from  
the  \mcatnlo\   calculation after inclusion of \PBM -TMDs (blue solid curves). 
It is interesting to observe that the contribution coming from the real hard partonic emission is small 
at low center-of-mass energies and at low \pt , but increases with increasing $\sqrt{s}$, thus allowing one 
 to study the contribution of multiple soft emissions in detail. 

\begin{figure}[h!tb]
\begin{center} 
\includegraphics[width=0.45\textwidth]{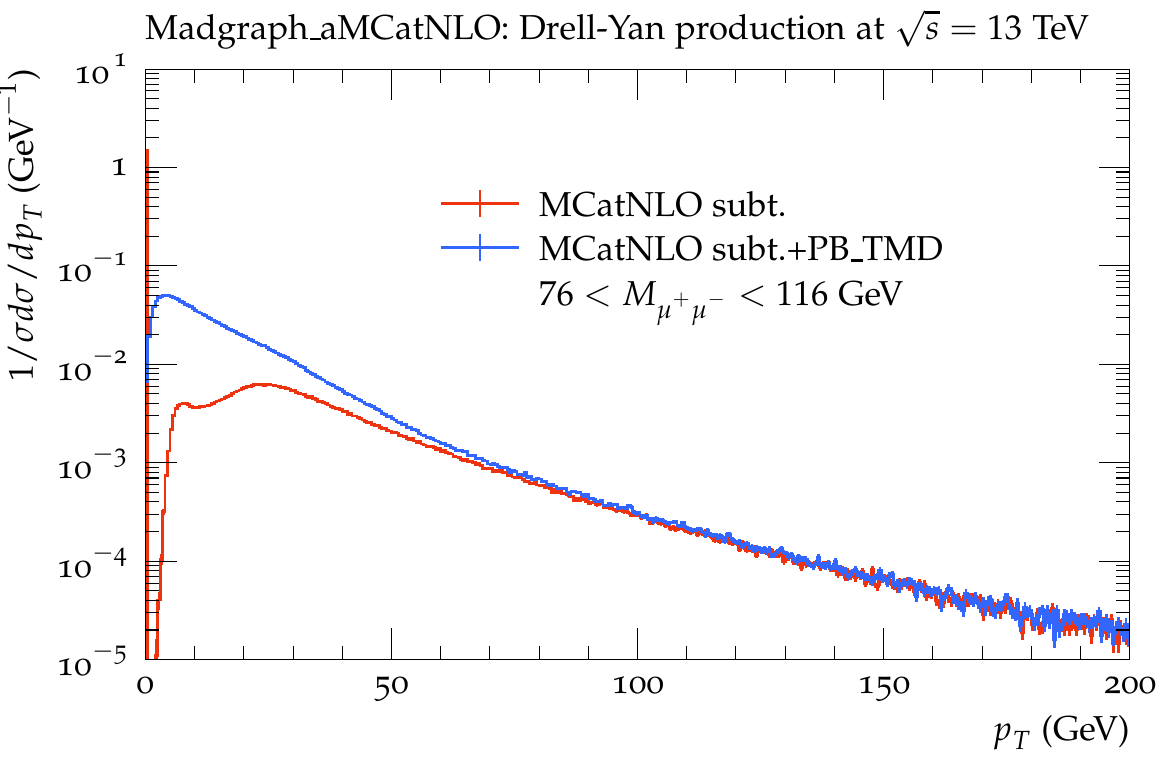} 
  \caption{\small Transverse momentum spectrum of \PZ\ production  at parton level (LHE level) for  subtraction terms and after inclusion of \PBM -TMDs at $\sqrt{s}=13$~\TeV. 
 }
\label{mcatnlo_lhe-LHC}
\end{center}
\end{figure}
\begin{tolerant}{1200}
In Fig.~\ref{mcatnlo_lhe-LHC} the distribution in transverse  momentum, with subtraction terms and after inclusion of \PBM -TMDs, is shown for LHC energies of $\sqrt{s}=13$~\TeV\  and for DY masses around the \PZ -mass.  At high $\sqrt{s}=13$~\TeV\  and at sufficiently large DY mass, the predictions with and without   \PBM -TMDs  become similar at large transverse momenta, supporting the simple expectation that for $\pt/\mdy \gsim 1 $   the transverse momentum spectrum is essentially driven by hard real emission.  
\end{tolerant}

\section{Low mass DY production}
\label{sec:lowmass}

We next apply the framework described in the previous section, based on the matching of   \PBM -TMDs  with  NLO, to the 
evaluation of DY  spectra at low DY masses. 

\subsection{Mass and transverse momentum spectra 
\label{sec:spectra}}
\begin{tolerant}{1200}
We start with the DY mass spectrum at low masses and low $\sqrt{s}$. In 
Fig.~\ref{DY_M} we present theoretical predictions obtained from \PBM -TMDs and NLO  matrix elements using \mcatnlo\ matching, 
and compare them with experimental measurements  
 for different center-of-mass energies  from  
 NuSea~\protect\cite{Webb:2003ps,Webb:2003bj}, R209~\protect\cite{Antreasyan:1981eg} and PHENIX~\protect\cite{Aidala:2018ajl}. 
We also show the theoretical uncertainties coming from the determination of the \PBM -TMDs as well as from the 
variation of the scale in the perturbative calculation. As already observed in Ref.~\cite{Martinez:2019mwt} for the case 
 of \PZ -production at the LHC, the contribution to uncertainties from the parton density turns out to be small compared to the one from the scale uncertainty.
Not included are the uncertainties coming from the variation of the intrinsic Gauss distribution ($q_s$), as this parameter was not constrained by the 
fits to HERA data~\cite{Martinez:2018jxt}. This  will be further discussed in Subsec.~\ref{q0_determiation}. 
\end{tolerant}

\begin{figure}[htb]
\begin{center} 
\includegraphics[width=0.35\textwidth]{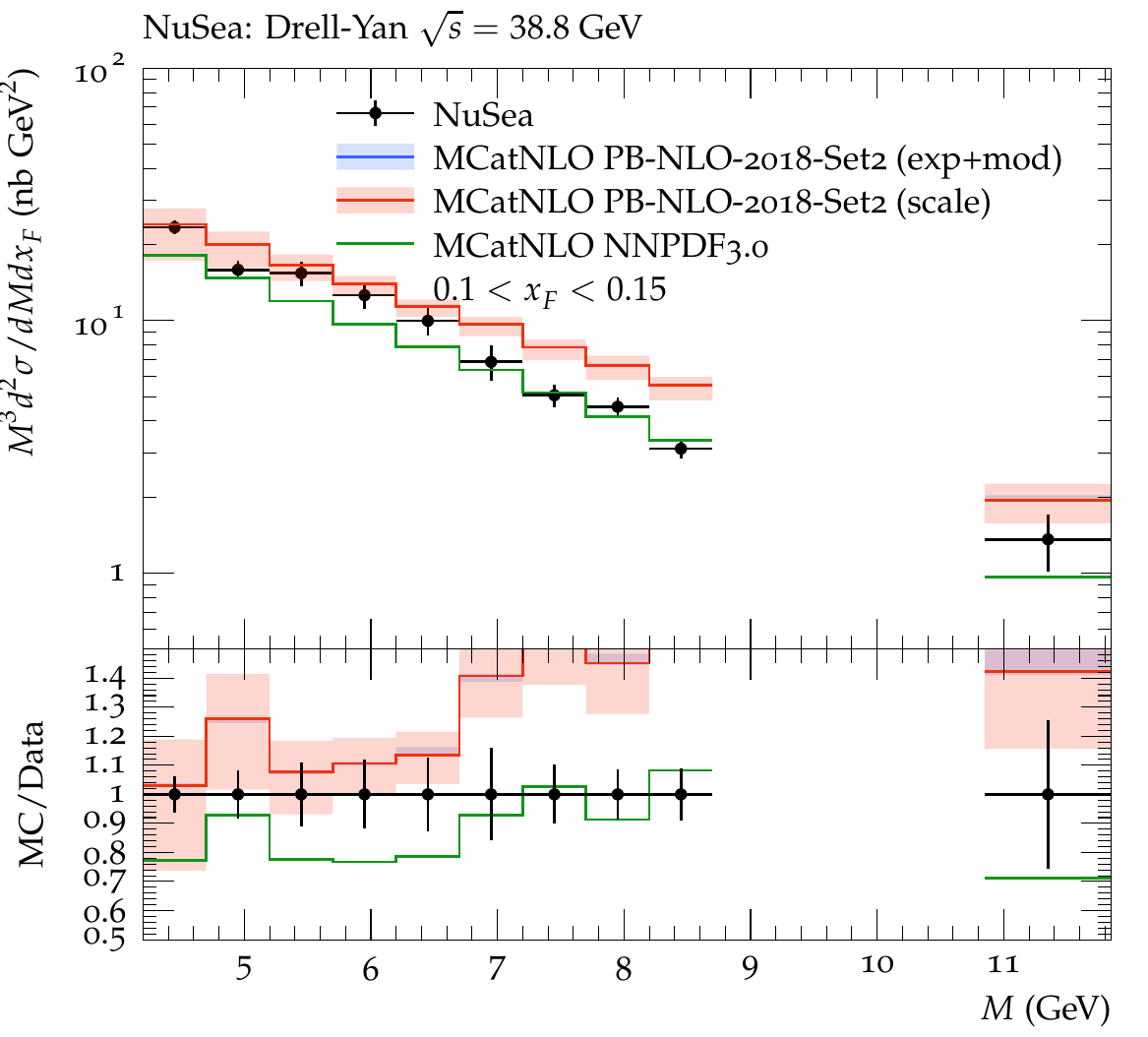} 
\includegraphics[width=0.35\textwidth]{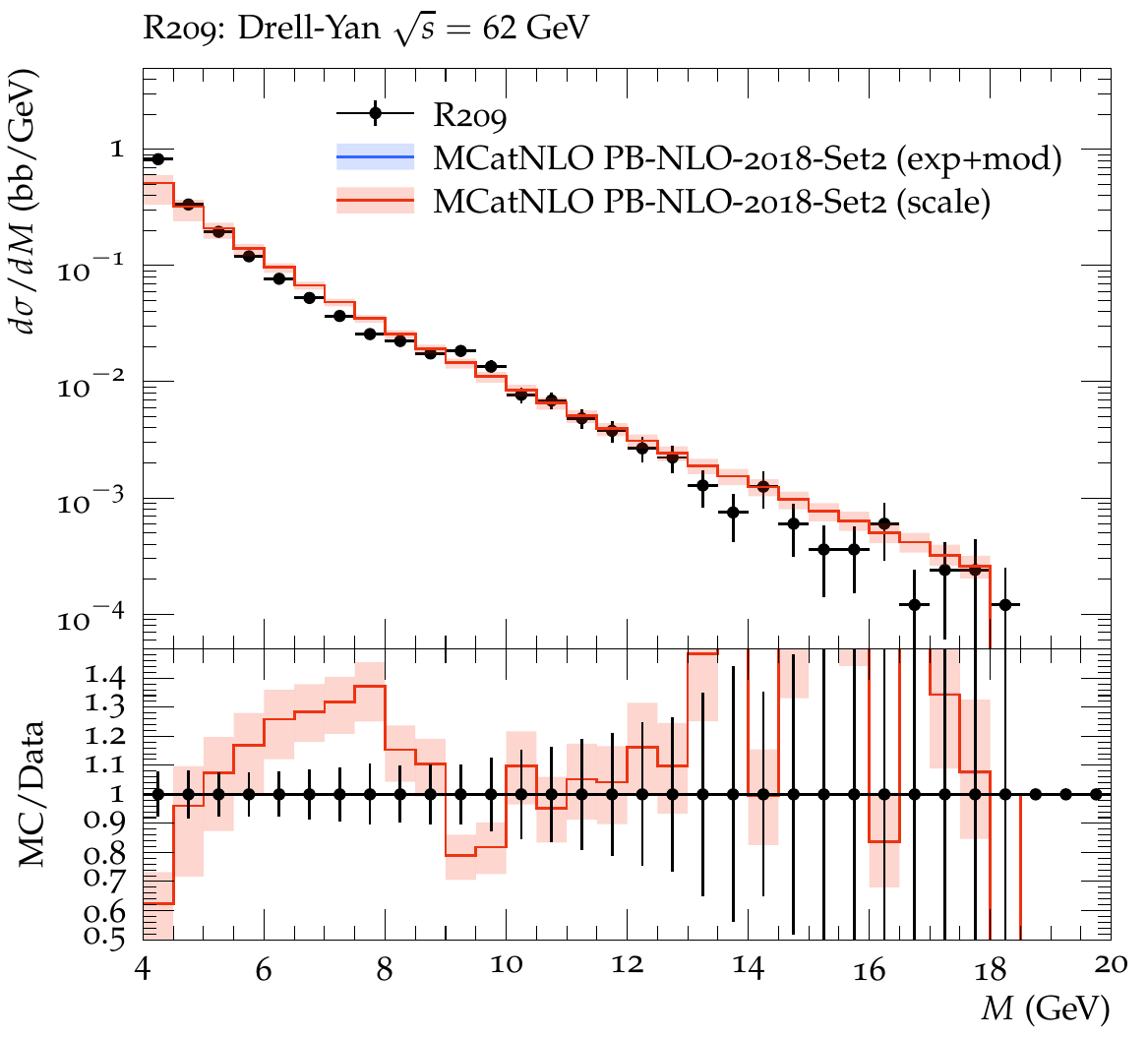} 
\includegraphics[width=0.35\textwidth]{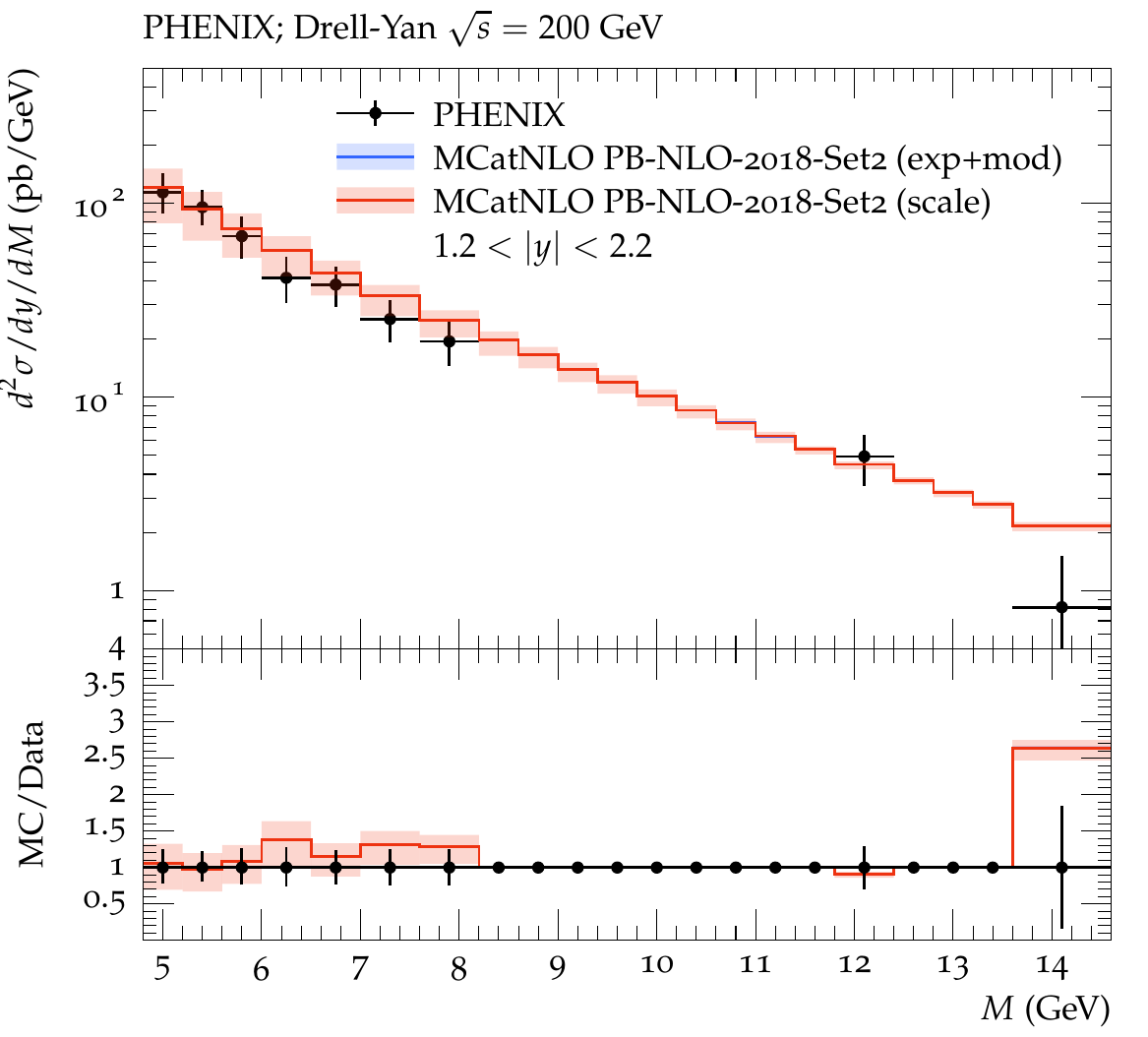} 
  \caption{\small Drell-Yan mass distribution production measured by NuSea~\protect\cite{Webb:2003ps,Webb:2003bj}, R209~\protect\cite{Antreasyan:1981eg} and PHENIX~\protect\cite{Aidala:2018ajl}  
   compared to predictions at NLO using \PBM -TMDs. For NuSea also the prediction using NNPDF3.0~\protect\cite{Ball:2014uwa} is shown.}
\label{DY_M}
\end{center}
\end{figure} 
The mass spectra in  Fig.~\ref{DY_M}  are generally well described by the \PBM -TMD + NLO calculation. 
  For the region of highest masses at lowest  $\sqrt{s}$ (NuSea experiment), we see in the top panel of Fig.~\ref{DY_M} that the description of experimental data 
  by the \PBM -TMD + NLO calculation deteriorates. This is  because we enter the large-$x$ region  where the 
  parton densities~\cite{Martinez:2018jxt} used in the calculation, which are determined from fits to  
   HERA data~\cite{Abramowicz:2015mha},  are poorly constrained. The description in this region can be readily improved 
   by using parton density sets from global fits. We show this in Fig.~\ref{DY_M} by plotting 
   the result from the  set NNPDF3.0~\cite{Ball:2014uwa}, obtained from global fits that include NuSea data~\cite{Webb:2003ps,Webb:2003bj}. 
On the other hand, 
 for the lowest mass region $\mdy <  6$ \GeV\  of NuSea the mass spectrum is well described.  We use this region to 
  investigate the transverse momentum spectrum.

In 
Fig.~\ref{DY_pt} we present theoretical predictions  from \PBM -TMDs and NLO matrix elements for transverse momentum spectra, and again we compare them 
with experimental measurements  
 for different center-of-mass energies  from  
 NuSea~\protect\cite{Webb:2003ps,Webb:2003bj}, R209~\protect\cite{Antreasyan:1981eg} and PHENIX~\protect\cite{Aidala:2018ajl}. 
The  \PBM -TMDs used in the calculation include an intrinsic (non-perturbative) transverse momentum spectrum 
parameterized as a Gauss distribution with width $ \sigma^2  =  q_s^2 / 2 $ (see eq.(\ref{TMD_A0})).  
\begin{figure}[h!tb]
\begin{center} 
\includegraphics[width=0.35\textwidth]{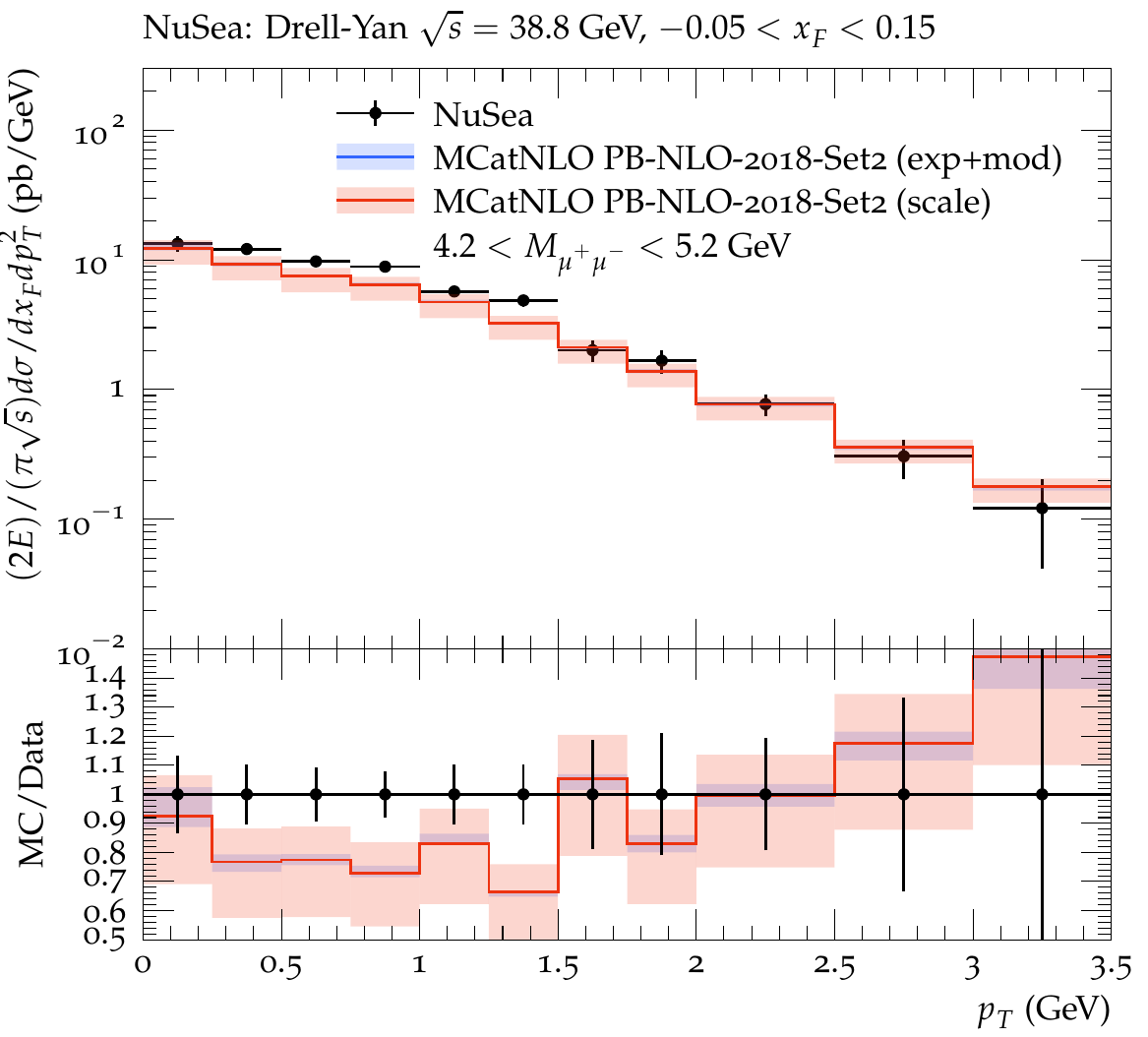} 
\includegraphics[width=0.35\textwidth]{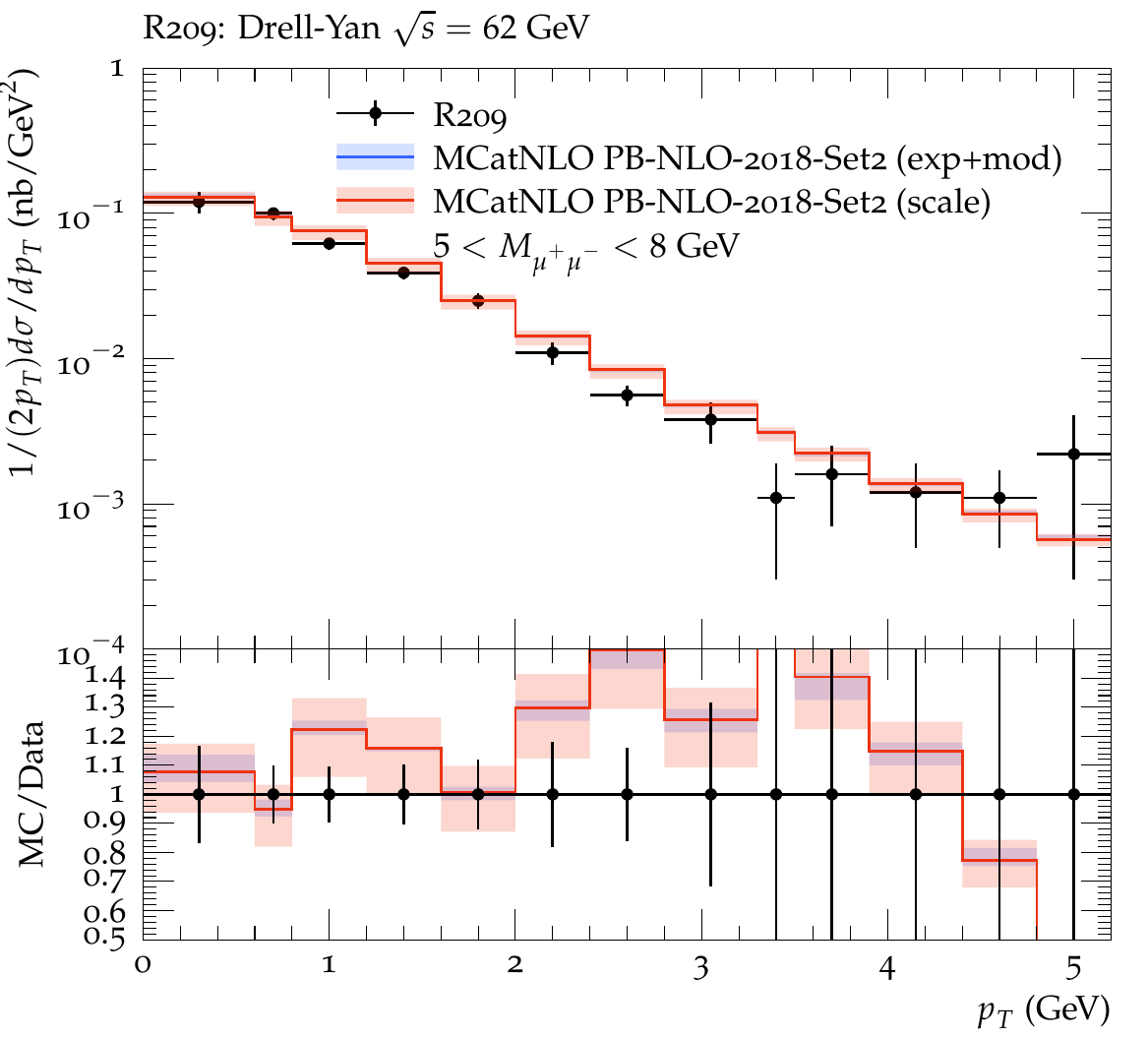} 
\includegraphics[width=0.35\textwidth]{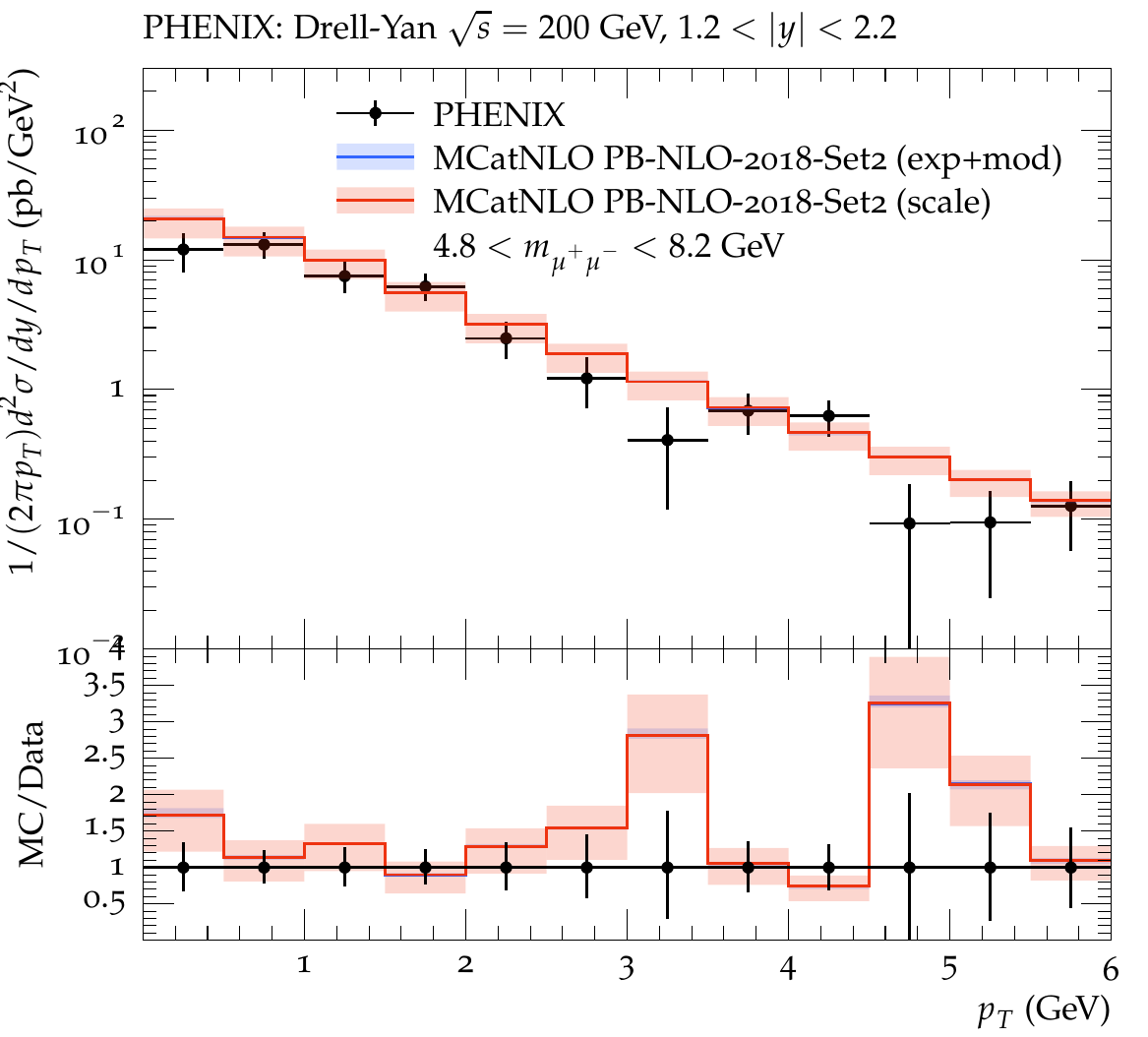} 
  \caption{\small Transverse momentum spectrum of Drell-Yan production measured by NuSea~\protect\cite{Webb:2003ps,Webb:2003bj}, R209~\protect\cite{Antreasyan:1981eg}, PHENIX~\protect\cite{Aidala:2018ajl}  compared to predictions at NLO using \PBM -TMDs. }
\label{DY_pt}
\end{center}
\end{figure} 
The quality of the description of the measurements (including independent variations of the factorization and renormalization scales by a factor of two up and down) is good with $\chi^2/ndf = 1.08, 1.27, 1.04$ for NuSea, R209 and PHENIX, respectively. The $\chi^2$ values are calculated using the full $\pt$ range. 
In the above discussion we have shown results for NuSea, R209 and PHENIX as representative of a broad range of different center-of-mass energies. We have 
obtained similar results for other data sets in this energy range such as E605~\cite{Moreno:1990sf}.  

In Fig.~\ref{Z_pt_CMS} we show the transverse momentum spectrum of \PZ -bosons at LHC energies of $\sqrt{s}=13$~\TeV\ as measured by CMS~\cite{Sirunyan:2019bzr} and compare it with predictions using  the same method of the above low-energy predictions and of Ref.~\cite{Martinez:2019mwt},  
 with the \PBM -TMD Set~2. We observe a very good description of the measurement (with $\chi^2/ndf = 0.8$ for $\pt < 80$ \GeV ). As discussed in 
 Ref.~\cite{Martinez:2019mwt}, the drop in the prediction at large transverse momenta comes from missing NLO contributions to \PZ + jet production, i.e., 
 ${\cal O} (\alpha_s^2)$ terms in the hard process calculation. 
\begin{figure}[h!tb]
\begin{center} 
\includegraphics[width=0.35\textwidth]{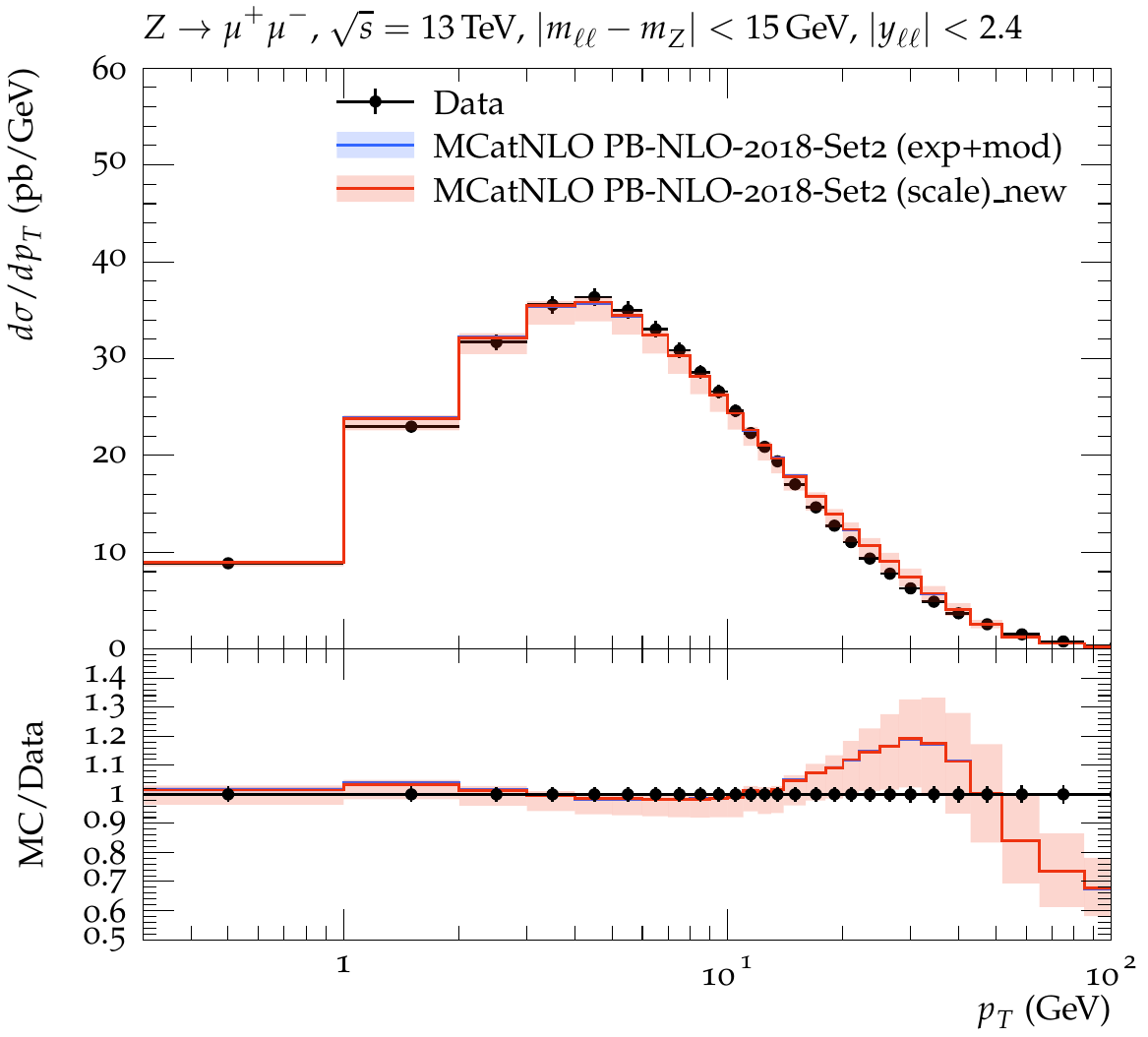} 
  \caption{\small Transverse momentum spectrum of \PZ\ production measured by CMS~\protect\cite{Sirunyan:2019bzr} compared to predictions at NLO using \PBM -TMDs. }
\label{Z_pt_CMS}
\end{center}
\end{figure}

\subsection{Determination of the non-perturbative (intrinsic) transverse momentum distribution
\label{q0_determiation}}
The low-mass DY measurements can be used to constrain the intrinsic transverse momentum distribution. In Fig.~\ref{chi2_gauss} we report the calculated $\chi^2/ndf$ as a function of $q_s$ obtained from the transverse momentum distributions of NuSea~\protect\cite{Webb:2003ps,Webb:2003bj}, R209~\protect\cite{Antreasyan:1981eg}, PHENIX~\protect\cite{Aidala:2018ajl}
(as shown in Fig.~\ref{DY_pt}). For the calculation of $\chi^2/ndf$ we use the full experimental uncertainties (except an overall normalization uncertainty) and the central values for the theory predictions (without inclusion of pdf and scale uncertainties, leading to a larger $\chi^2/ndf$ as the one reported in the previous subsection).
\begin{figure}[h!tb]
\begin{center} 
\includegraphics[width=0.45\textwidth]{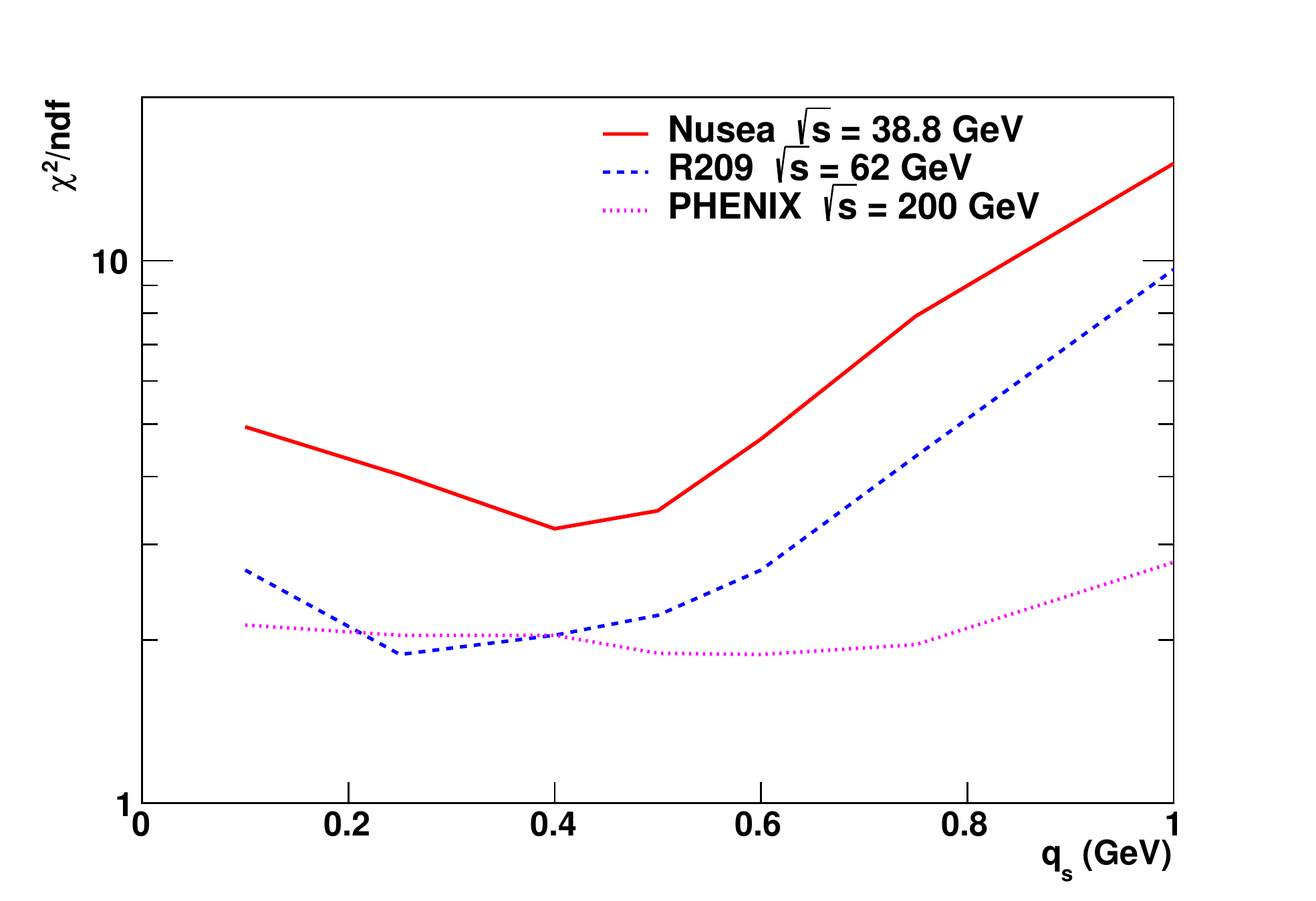} 
  \caption{\small The $\chi^2/ndf$ as a function of the width of the intrinsic transverse momentum distribution, obtained from a comparison of the measurements (NuSea~\protect\cite{Webb:2003ps,Webb:2003bj}, R209~\protect\cite{Antreasyan:1981eg}, PHENIX~\protect\cite{Aidala:2018ajl}) with a prediction at NLO using \PBM -TMDs. 
  For the theory prediction only the central value is taken, but no uncertainty from scale variation is included.}
\label{chi2_gauss}
\end{center}
\end{figure} 

A clear minimum is found for NuSea and R209 measurements, with values  of $q_s \sim 0.3 - 0.4 $ \GeV . On the other hand, 
 the PHENIX measurement shows 
 little sensitivity to the choice of $q_s$, which is understandable since only two values for $\pt <1$ \GeV\ are measured, while the other experiments have  a finer binning. 
It is interesting to note that the values of intrinsic transverse momentum  determined from low-mass DY 
are rather close to the value of $q_s=0.5$ \GeV\   that was assumed in \PBM -Set2 \cite{Martinez:2018jxt}, determined from fits to inclusive DIS data from HERA which are 
not sensitive to  intrinsic-\kt .

\subsection{Comments on the low-mass region 
\label{comments}}

It has been observed in~\cite{Bacchetta:2019tcu}   that perturbative fixed-order  calculations at ${\cal O} (\alpha_s)$ and 
${\cal O} (\alpha_s^2)$  
 in collinear factorization are not able to describe 
the  measurements of DY transverse momentum spectra at fixed-target experiments   in the region $\pt/\mdy \sim 1$.     
We remark that this is consistent with the observation which we have made in  Fig.~\ref{mcatnlo_lhe} 
 that, in this kinematic region, the contribution from the real hard emission is small  compared to the contribution from multiple parton radiation,  
embodied in the \PBM -TMD evolution. Indeed, Fig.~\ref{mcatnlo_lhe}  indicates that a purely collinear   NLO  calculation 
would not give a realistic description of the DY spectrum for $\pt/\mdy \sim 1$ at low energies. On the other hand, 
Fig.~\ref{mcatnlo_lhe-LHC}  illustrates that the situation is very different at the LHC:  in the region around the Z mass 
shown in Fig.~\ref{mcatnlo_lhe-LHC},   hard real emission dominates the transverse momentum spectrum  for $\pt/\mdy \sim 1$, so that a
purely collinear   NLO  calculation gives a good approximation to the DY process  for $\pt/\mdy \sim 1$ at the LHC. 

The comparison of theoretical predictions with  transverse momentum  measurements from NuSea~\cite{Webb:2003ps,Webb:2003bj}    
in the top  panel of Fig.~\ref{DY_pt} confirms that the inclusion of multiple parton emissions, taken into account by the 
  \PBM -TMD evolution equation~\cite{Hautmann:2017fcj} (see also discussion in~\cite{Hautmann:2019biw}),    
is essential to describe the region $\pt/\mdy \sim 1$ at low energies.  This physical picture is supported by the comparison of 
theoretical predictions with measurements at the increasingly high energies of R209~\cite{Antreasyan:1981eg} and 
PHENIX~\cite{Aidala:2018ajl} in the middle and bottom panels of Fig.~\ref{DY_pt}.  
Going up to LHC energies in Fig.~\ref{Z_pt_CMS}, we see that the \PBM -TMD + NLO calculation 
describes  the  spectrum well all the way up to transverse momenta  $\pt \sim \mdy $ (while for even  higher \pt\  a deficit is observed due to 
 the missing DY + jet NLO correction --- see discussion in~\cite{Martinez:2019mwt}).  
 
 Our calculation thus indicates that  at low energies QCD contributions beyond fixed order (${\cal O} (\alpha_s)$,  
 ${\cal O} (\alpha_s^2)$, etc.)   are important to describe the region $\pt/\mdy \sim 1$, unlike the case of LHC energies where 
fixed order calculations are sufficient to describe the region  $\pt/m_Z \sim 1$. We have taken into account  all-order 
contributions through the  \PBM -TMD  evolution formalism, and found that this allows one to describe well 
the transverse momentum spectra.  

To sum up,  the DY transverse momentum in the low-mass region is sensitive to  both  finite-order QCD contributions and all-order QCD multi-parton radiation.  
Theoretical predictions depend on the matching procedure between these contributions. 
Once this is accomplished, low-mass DY measurements are well described 
and can provide a wealth of information on non-perturbative QCD dynamics. 
In this paper the matching is performed, in the spirit of~\cite{Collins:2000gd}, with  \PBM -TMDs and \mcatnlo\  (alternative methods of  
 matching are e.g.~those inspired by~\cite{Collins:1984kg}).

\section{Discussion}
\label{sec:disc}

To put the results of this work in a broader context, one may start from a simple scenario in 
which one hopes to describe high-${\ensuremath{p_{\rm T}}}$ dynamics by perturbative NLO calculations combined 
with collinear parton densities,  and low-${\ensuremath{p_{\rm T}}}$ dynamics by non-perturbative TMDs based, in 
the simplest model, on intrinsic-${\ensuremath{k_{\rm T}}}$ Gauss distributions.  One may wonder whether these 
two elements, NLO collinear calculations for perturbative high-${\ensuremath{p_{\rm T}}}$ physics and 
intrinsic-${\ensuremath{k_{\rm T}}}$ 
TMD distributions for nonperturbative low-${\ensuremath{p_{\rm T}}}$ physics, are sufficient to provide a satisfactory 
description of the transverse momentum spectrum over all kinematic regions. The analysis of  
this paper illustrates that this simple approach cannot be guaranteed to give the correct physical 
picture in all phase space configurations. The key element which is missing in this simple approach 
is QCD multiple-parton radiation, and the analysis of this paper shows that (predominantly infrared)  
components of this radiation become essential in the region ${\ensuremath{p_{\rm T}}} 
 \sim 1 -  10$ GeV $ \sim {\cal O} ( {\ensuremath{m_{\rm DY}}} ) $ of low-energy DY experiments. 
It also shows, more specifically, 
that such effects are essential for the transverse momentum spectrum, while 
they do not influence very much the mass spectrum integrated over transverse momenta.  

If  such contributions are to be included, one could imagine doing this in different manners. In this work 
we have done this by the PB method. This may be regarded as being  well-suited to this problem,  
because i) it includes multiple-parton radiation through the evolution of TMDs, ii)    it incorporates the 
intrinsic-${\ensuremath{k_{\rm T}}}$ distribution  as a nonperturbative 
boundary condition to  a well-defined branching evolution equation in terms of perturbatively calculable  kernels, 
and iii)  it is matched through  \mcatnlo\  to NLO  hard-scattering functions. 
It thus contains the three main inputs which are  essential to  the physical picture described above. In particular, 
the  kernels describing multi-parton  radiation through TMD evolution are given in terms of  
Sudakov form factors, real-emission splitting 
functions, and  angular-ordering phase space constraints, which are important to correctly take into account 
infrared gluon emission. 

The analysis performed in this paper leads to different conclusions from those which have appeared in the literature   pointing  
to difficulties~\cite{Bacchetta:2019tcu} in describing the 
low-mass and low-energy DY measurements and to the  ``$\qt$ crisis'' scenario~\cite{Aidala:2020mzt,correlations:2020feb}.  
The analysis in this paper indicates that,  provided infrared multi-parton radiation 
is included (e.g., through \PBM -TMD evolution),  theoretical predictions describe 
low-mass and low-energy DY measurements well. It further shows 
that such measurements provide enhanced  sensitivity to 
intrinsic ${\ensuremath{k_{\rm T}}}$ compared to the case of 
high-energy experiments. They 
can thus be usefully exploited for  determinations of nonperturbative TMDs. 
The analysis in this paper  
 also stresses the difference between the behavior discussed above for  the  region ${\ensuremath{p_{\rm T}}} \sim 
{\ensuremath{m_{\rm DY}}} $ of low-energy DY experiments and the behavior in the  region measured at the LHC with 
${\ensuremath{p_{\rm T}}} \sim {\ensuremath{m_{\rm DY}}} \sim 100$ GeV. In the latter, no large correction is expected to 
purely-collinear finite-order perturbative calculations. This confirms  that 
arguments purely based 
on scaling  in the ratio ${\ensuremath{m_{\rm DY}}}  /  {\ensuremath{p_{\rm T}}} $ are not sufficient,   
due to both the running of the strong coupling, and the   role of infrared emission.

Other approaches would be possible as well.  For instance, parton showers take into account multiple parton radiation in a manner 
alternative to the  \PBM -TMD method. They can be matched to NLO matrix elements. Most parton shower Monte Carlo 
also model intrinsic-${\ensuremath{k_{\rm T}}}$ effects. In this respect, it is noteworthy that 
the  \herwig\  study~\cite{Gieseke:2007ad}  found good agreement with DY measurements at low energy, provided 
parameters for the parton shower and intrinsic   ${\ensuremath{k_{\rm T}}}$  were suitably tuned, and it 
 should be interesting to also reanalyze this in \pythia\  and other Monte Carlo generators.  
 The agreement with DY data found 
 in~\cite{Gieseke:2007ad}  underlines the relevance of infrared multiple emissions 
 (taken into account, in this calculation, by showering)  
 for the DY region of the low-energy experiments. 
 
 However, significant 
 differences exist between the parton shower approach and the 
 \PBM -TMD approach. One significant difference   is that in 
the    \PBM -TMD method  nonperturbative TMD densities are defined and determined from fits to 
experimental data, which places constraints on fixed-scale inputs to evolution, while 
 in  parton showers  the   parton densities are not used to constrain evolution, and instead 
  nonperturbative physics parameters are tuned. This may have an impact on the size of 
  intrinsic-${\ensuremath{k_{\rm T}}}$ effects in the two approaches. On one hand, 
    in the case of the PB method we have 
  seen in this work that intrinsic ${\ensuremath{k_{\rm T}}} \simeq q_s / \sqrt{2} $ with 
  $q_s \in ( 300 ,  500) $  MeV provides predictions which describe well DY measurements 
  across the energy range from NuSea $\sqrt{s}=38.8$ GeV to the LHC $\sqrt{s}=13$   
   TeV. On the other hand, to our knowledge 
    it is not yet clear at present whether tuning of parton shower 
   generators to LHC and low-energy data would result in similarly mild 
   ${s}$-dependence of the intrinsic ${\ensuremath{k_{\rm T}}}$, or whether it would require a much 
   stronger  ${s}$-dependence. 
   
   A further  significant  difference 
   between the shower and  \PBM -TMD approaches 
   is that in the shower calculation~\cite{Gieseke:2007ad}   
   the showering scale is lowered,  with respect to the case of  the LHC,  to describe the 
   low-energy region. In contrast, in the     \PBM -TMD calculation of this paper 
    the initial evolution scale is not changed, 
   and the same  starting scale $\mu_0 \simeq 1 $ GeV is applied for the LHC and for the lower-energy 
NuSea, R209 and PHENIX experiments.    
       We think that the investigation of these differences and their interpretation  will be  
       important questions to be examined, particularly  to elucidate contributions from low-momentum regions. 

Another possible approach is based on analytic CSS~\cite{Collins:1984kg} resummation. 
In this formalism too the contributions from multiple soft-gluon emission, intrinsic ${\ensuremath{k_{\rm T}}}$ and 
NLO hard-scattering functions can be included. The formulation is however very different from that in the PB method. 
In particular, the matching procedure~\cite{Collins:1984kg} (involving the so-called $W$ and $Y$ terms)  differs 
from the matching used in this paper, which is of the type studied in~\cite{Collins:2000gd}. Also the way to include 
 intrinsic transverse momentum effects (in ${\bf b}$ or ${\ensuremath{k_{\rm T}}}$ space) differs between CSS and PB. 
We expect the region ${\ensuremath{p_{\rm T}}}  \sim 1 -  10$ GeV of low-energy DY experiments to be particularly sensitive to  
 such  differences in the matching and intrinsic ${\ensuremath{k_{\rm T}}}$ effects. 
We therefore think that much is to be learnt from a detailed comparison in this region.

\section{Conclusion}
\label{sec:concl} 

We have investigated the transverse momentum spectra of DY lepton-pair production at small DY masses and low center-of-mass energies by matching  \PBM -TMD distributions to NLO calculations via \mcatnlo . We use the same \PBM -TMDs and \mcatnlo\ calculations as we have used  for \PZ -production at LHC energies in Ref.~\cite{Martinez:2019mwt}. We observe a very good description of the measurements by the NuSea collaboration at $\sqrt{s}=38$~\GeV , R209 at $\sqrt{s}=62$~\GeV\ and PHENIX at $\sqrt{s}=200$~\GeV , with values of $\chi^2/ndf \sim 1$ for all measurements. 
We use the low-mass DY  measurements to determine the best value for the width of the intrinsic Gauss distribution, and find a value of $q_s \sim 0.3 - 0.4 $ \GeV , slightly smaller than $q_s=0.5$ \GeV\ used in the \PBM -TMD Set~2 distributions~\cite{Martinez:2018jxt}.

The very good description of low-mass DY measurements is achieved by a combination of a collinear NLO calculation (including the appropriate subtraction terms to avoid double counting) with the \PBM -TMDs. We find that,  at low DY mass and low $\sqrt{s}$,  even in the region of $\pt /\mdy  \sim 1$ the contribution of 
 QCD multi-parton radiation (included in the evolution of  \PBM -TMDs in terms of Sudakov form factors, resolvable splitting functions and 
phase space constraints) is essential to describe the measurements, while at larger masses  ($\mdy \sim m_{\PZ }$) and  LHC energies this  
contribution is small  in the region of  $\pt /\mdy  \sim 1$.

The results which we have presented in Figs.~\ref{mcatnlo_lhe} and~\ref{DY_pt}, in particular, provide a 
new perspective on the ``$\qt$ crisis''  recently discussed in the literature (see e.g. contributions 
 in Refs.~\cite{Bacchetta:2019tcu,Aidala:2020mzt,correlations:2020feb})  with regard to 
measurements of transverse momentum spectra at low mass.   
Fig.~\ref{mcatnlo_lhe} illustrates that,  in the  kinematic region  $\pt/\mdy \sim 1$ of experiments at low center-of-mass energies $\sqrt{s}$, 
hard real emission does not dominate the transverse momentum spectrum, in contrast to the case of the analogous kinematic region 
around the Z boson mass at the LHC.  Correspondingly, NLO collinear calculations are not sufficient 
to describe the region of  low-energy DY measurements and multi-parton radiation contributions need to be taken into account.   
On the other hand, Fig.~\ref{DY_pt} shows that once the matching of NLO and multi-parton contributions is accomplished, as is done in the 
present study using  the \PBM -TMD  formalism,   low-mass DY measurements can be well described 
over a broad range of center-of-mass energies  $\sqrt{s}$ including the NuSea, R209 and PHENIX experiments. 
The matching in the present paper is carried out via  \PBM -TMDs and \mcatnlo\  with an approach similar to~\cite{Collins:2000gd}.   
Low-mass DY data can thus be exploited to  extract information on non-perturbative TMD dynamics.

\vskip 0.3 cm 

\noindent 
{\bf Acknowledgments.}
We thank E.~Aschenauer, A.~Bacchetta, V.~Bertone, A.~Bressan, M.~Diefenthaler, G.~Ferrera, B.~Parsamyan, G.~Schnell   
 and A.~Vladimirov for  useful discussions. 
We thank Yue Hang Leung for discussions on the PHENIX measurement. We thank A. Siodmok for pointing us to their study on \herwig .
FH acknowledges the support and hospitality of  DESY, Hamburg  while part of this work was being done. 
STM thanks the Humboldt Foundation for the Georg Forster research fellowship  and 
gratefully acknowledges support from IPM.    
QW and HY acknowledge the support by the Ministry of Science and Technology under grant No. 2018YFA040390 and 
by the National Natural Science Foundation of China under grant No. 11661141008.

\providecommand{\href}[2]{#2}\begingroup\raggedright\endgroup

\end{document}